# Giant Enhancement in Ferroelectric Polarization under Illumination

Hitesh Borkar[1,2], Vaibhav Rao[1,3], M Tomar[4], Vinay Gupta[5], J. F. Scott[6], Ashok Kumar[1,2,]*

[1]CSIR-National Physical Laboratory, Dr. K. S. Krishnan Marg, New Delhi 110012, India
[2]Academy of Scientific and Innovative Research (AcSIR), CSIR-National Physical Laboratory (CSIR-NPL) Campus, Dr. K. S. Krishnan Road, New Delhi 110012, India
[3]Solar and Alternative Energy, Amity University, Jaipur, Rajasthan, 302006, India
[4]Department of Physics, Miranda House, University of Delhi, Delhi 110007, India
[5]Department of Physics and Astrophysics, University of Delhi, Delhi 110007, India
[6]Department of Chemistry and Department of Physics, University of St. Andrews, St. Andrews KY16 ST, United Kingdom

## Abstract:

We report optical enhancement in polarization and dielectric constant near room temperature in $Pb_{0.6}Li_{0.2}Bi_{0.2}Zr_{0.2}Ti_{0.8}O_3$ (PLBZT) electro-ceramics; these are doubly substituted members of the most important commercial ferroelectric $PbZr_{0.2}Ti_{0.8}O_3$ (PZT:20/80). Partial (40%) substitution of equal amounts of $Li^{+1}$ and $Bi^{+3}$ in PZT: 20/80 retains the PZT tetragonal structure with space group *P4mm*. Under illumination of white light and weak 405-nm near-ultraviolet laser light (30 mW), an unexpectedly large (200-300%) change in polarization and displacement current was observed. Light also changes the dc conduction current density by one to two orders of magnitude with a large switchable open circuit voltage ($V_{oc}$ ~ 2 V) and short circuit current ($J_{sc}$ ~ $5\times10^{-8}$ A). The samples show a photo-current ON/OFF ratio of order 6:1 under illumination of weak light.

*Corresponding Author: Dr. Ashok Kumar (Email: ashok553@nplindia.org)

1. **Introduction:**

Almost fifty years ago the functionalities of optically sensitive ferroelectrics were thoroughly discussed by Fridkin.[1] He was among the first to recognize explicitly that many ferroelectrics were semiconductors and not "insulators," as they were usually termed at that time. Although he emphasized relatively narrow-gap materials such as SbSI, it is now recognized that the majority of oxide ferroelectrics have bandgaps of order 3.5 eV, not much different from wide-gap II-VI's such as ZnO or III-V's such as GaN. Many ferroelectrics exhibit photo-ferroelectric phenomena due to strong influence of non-equilibrium charge carriers under illumination. These phenomena are useful for several applications, including photo-driven actuators, highly sensitive photo-detectors, low power high density nonvolatile memory elements, and photovoltaic devices.

In general when light is incident on an electro-ceramic with a wavelength corresponding to the energy bandgap of the system, the incident photons generate charge carriers and develop a polarity shift.[1,2,3,4] The illumination of light on photon-sensitive polar ferroelectric materials enhances photo-conductivity, increases or decreases polarization, shifts ferroelectric phase transition temperatures, causes photo-striction (e.g., in SbSI), and may modify domain structure at the nanoscale level. These phenomena are often lumped generically under the label "photo-ferroelectricity" and have been extensively studied.[1] These effects can be employed for novel microelectronic applications using ferroelectric thin films as nonvolatile memory elements with electrical WRITE and optical READ logic and as micro-electromechanical (MEMS) systems with optical detectors.[5]

The effects of light on fatigue, poling, polarization, retention and conductivity of ferroelectric materials has been widely studied in several ferroelectric materials.[6,7,8,9,10,11,12,13,14] It is known that functional oxides are promising materials due to their robust and relatively inert physical properties and potential applications[15,16,17] Among oxides, ferroelectric materials exhibit excellent functionalities for such device embodiments, due to presence of spontaneous polarization.[18,19] Here we consider the effects of light on ferroelectric polarization, dielectric constant, loss tangent and conductivity. Under illumination conditions, light orients the polarization in a particular direction parallel or antiparallel to the direction of applied electric field and also generates electronic polarization regions in the crystal.[3,20] The ferroelectric photovoltaic effects, including change in capacitance and dimensions under intense light, are attributed directly to the charge carriers and to the spontaneous polarization.[21,22] Researchers have also developed several optically active ceramics capable of storing optical information.[23,24,25,26] The present study is devoted to light-assisted change in polarization and dielectric properties of polar materials.[27,28] These changes can be large; photo-stimulated ferroelectric phase transitions have been studied that suggest a strong interaction between the electron and lattice sub-systems near the Curie temperature.[1,29,30,31,32]

There are a vast number of reports in the literature on A- and B-site substitution of PZT-based systems for improving their functional properties which remarkably enhanced the optical, ferroelectric, dielectric and leakage-current properties by modification of the surface morphology and the chemical distribution of their elements. Such A- and B-site substitution significantly alters the ferroelectric phase transition temperature and functional properties. Here we report the effects of illumination of light on PZT with 40% substitution of $Pb^{+2}$ by equal amounts of $Bi^{+3}$ and $Li^{+1}$ cations, thus maintaining charge neutrality. This composition was optimized after analysis of physical properties of a wide range of Li and Bi doping in the PZT specimens.

## 2. Experimental Methods and Techniques:

"Green" electro-ceramic powders of $(Pb_{0.6}Li_{0.2}Bi_{0.2})(Zr_{0.2}Ti_{0.8})O_3$ (PLBZT) were synthesized by a solid-state reaction route. The starting precursors of PbO, $Li_2CO_3$, $Bi_2O_3$, $ZrO_2$, and $TiO_2$ (from Alfa Aesar, purity$\leq$ 99.9%) were weighed in their proper stoichiometric proportion and mechanically ground in agate mortar for about 2 h in the presence of IPA (isopropyl alcohol). The physically mixed powder was dried before calcination at 850°C for 10 h with ramp heating at 5 degrees per minute. The calcined powder was thoroughly mixed with a binder PVA (polyvinyl alcohol) for pelletization. The calcined powder was pressed uniaxially at 5–6 tons per square inch into 10 mm diameter pellets with 1-mm thickness. These pellets were further sintered at 1200 °C for 4 h to achieve high density (ca. 95% of the theoretical maximum density). An extra 10% of PbO, $Li_2CO_3$, and $Bi_2O_3$ were added in stoichiometric amounts to compensate our measured (XRD and XPS) empirical loss during the high temperature calcination and sintering process.

Powder X-ray diffraction (XRD) using Cu-$k_\alpha$ ($k_\alpha$=1.545A°) monochromatic radiation for 2θ between 20° and 60° was utilized. XRD data were refined via Rietveld refinement technique using the Fullprof suite. Thin pellets ~300-500 μm were made by mechanical polishing for electrical characterization. To observe the effect of light on electrical properties, monochromatic laser light having wavelength 405 nm and a white light source having energies 30 mW/cm$^2$ and 60 mW/cm$^2$ respectively were used to generate the carriers. The polarization-electric field (P-E) hysteresis loops and leakage current with and without light were obtained with an indium tin oxide (ITO) transparent electrode, using a Radiant Ferroelectric Tester. For this transparent conductive electrode, an ITO film using rf-magnetron sputtering technique was deposited on one side and silver paint on the other. Temperature dependent dielectric properties were studied at various frequencies (100 Hz to 1MHz) in the temperature range from 25 °C to 450 °C using an LCR meter (HIOKI-3532-50) at an amplitude of 0.5 V. A high-resolution transmission electron microscope (HRTEM), Technai G20 at 300 kV, was used.

## 3. Results and Discussion:

### 3.1. Crystal Structure and Microstructures

The X-ray diffraction patterns of PLBZT ceramics are shown in Fig.1 to have tetragonal crystal structure with *P4mm* space group symmetry. The diffraction peaks assigned correspond to tetragonal PZT crystal structure, matched to JCPDS file number 33-0784. The average ionic size of $Li^{+1}$ and $Bi^{3+}$ ions nearly match that of $Pb^{2+}$ ions, which together with the average charge favors substitution at the Pb-site. The Rietveld analysis, refinement parameters and Wycoff positions are listed in Tables 1 and 2 of the Supplementary Information. The data imply a tetragonal crystal structure having *P4mm* space group symmetry and lattice parameters *a=b=3.929(9) Å* and *c=4.124(5) Å* with preferred orientation of lattice structure in direction of (101) planes. A HRTEM micrograph is given in the inset of Fig.1 that confirms that the (100) lattice plane matched the tetragonal structure obtained from these refinement parameters. The TEM image of large PLBZT nano-size ceramic particles/crystallites is given in Supplementary Fig.1. The chemical compositions of PLBZT pellets used for electrical measurements were mapped from large area SEM-EDX analysis; the experimental composition values matched the initial compositions within the equipment uncertainty ± 10%, as shown in Supplementary Table 3. Lithium is a light element, and it is not possible to detect it from EDX, which also adds extra errors in calculation. The Rietveld analysis parameters for PLBZT ceramics are *$\chi^2$ =1.85,*

*Rp=19.0, Rw=19.02,* and *Rexp=14.11*, which verifies our inference for the crystal structure within the fitting errors.

### 3.2. Dielectric Spectroscopy

Wide-range temperature- and frequency-dependent dielectric measurements were performed in the dark and under 405-nm monochromatic light, as shown in Fig. 2(a-f). The dielectric data indicate a small enhancement in dielectric constant and loss tangent at room temperature under laser illumination compared with dark measurements. These changes are around 3-4 % ($\varepsilon_{light}$-$\varepsilon_{dark}$)/($\varepsilon_{dark}$) in dielectric constant and loss in the low frequency (< 10 kHz) regimes at room temperature (Figs. 2(a.b)); however, the effect was negligible for higher probe frequency. Room-temperature enhancement in dielectric properties is thought to be due to "super-bandgap excitation" under illumination.[33] The changes in dielectric data are comparable to previously reported results on ferroelectric liquid crystals.[34,35] Temperature- dependent dielectric constant and loss tangent data at 1 kHz and 5 kHz are shown in Fig. 2 (c-f) for clarity; however, a complete range of dielectric and tangent loss data (dark and illuminated) is provided in supplementary Fig.2. As depicted in Fig. 2 (c,e) the ferroelectric phase transition temperature ($T_c$) of PLBZT under dark conditions occurred around 380 °C, comparable with that known for ferroelectric PZT (20/80). However, the magnitude of dielectric constant is significantly suppressed near the ferroelectric phase transition temperature (i.e. percentage suppression $= \frac{\varepsilon_{dark} - \varepsilon_{illu}}{\varepsilon_{dark}} X100$ is nearly 63% at 1 kHz under illumination); the effects were less prominent with increase in probe frequency. Interestingly, the magnitude of loss tangent and magnitude of $T_c$ also decrease under illumination at elevated temperature. Illumination significantly alters the nature of the ferroelectric phase transition; the sharp (in temperature) phase transition becomes a diffuse phase transition. These results indicate that at room temperature illumination enhances the dielectric constant and loss tangent; however, the effect was large but opposite in sign near the phase transition temperature (at elevated temperatures). These results indicate a direct connection between the charge carriers, temperature, illumination of light, and the structural phase transition. Such effects may be opposite in sign if the optical excitation of deep traps in these ferroelectrics behaves in the opposite way from that of carriers very near the bandgap.[1] The present experiments do not clearly discriminate between thermal effects and effects due to optical generation of increased carrier densities. Since the system is in steady state but not thermal equilibrium, free-energy models are not very useful.

### 3.2. Polarization, Displacement current and PUND study

Polarization-electric-field (P-E) loops of PLBZT were measured under dark and illumination conditions in the range of E-field ±80 kV/cm at 2 Hz and 5 Hz probe frequencies. P-E loops of bulk electro-ceramics do not respond well for higher probe frequencies.[36] Fig. 3 shows P-E loops in the dark and under illumination of white light and 405 nm monochromatic laser light for unpoled (a,b) and poled (c,d) samples at 2 and 5 Hz probe frequencies. A large change in remanent polarization was observed in the range of 100-250% under 30 mW 405 nm wavelength laser light and in the range of 50-150% under 40 mW white light, depending on probe frequencies, electrical field, and light intensity. These effects significantly decreased after poling under an electric field of E = 20 kV/cm for 2 hours (h). In general field-poling improves the polarization due to alignment of domains in the direction of the E-field and neutralizes the mobile charge carriers which may have accumulated across the grain boundaries and near defects. After E-poling, the free charge carriers were effectively neutralized and hence provide less contribution to the leakage current and polarization (an electret effect). Electronic polarizations can contribute to the overall polarization, which may responsible for "enhancement in polarization" under illumination. An excellent discussion has been given by Dawber *et al.*[4] With regard to the possible origin and direction of electronic polarization in perovskite $BaTiO_3$, they suggest that it is very likely that electronic polarization significantly affects the ionic polarization due to displacement of central cations and 2p electron orbitals for oxygen ions. The contribution of electronic polarization (compared with ionic) is not known in the present work, but the role of Li/Bi substitutioin for Pb suggests it is large. Recently similar results were obtained in molecular ferroelectric crystals, which showed that electronic polarization can be twenty times greater than that of ionic polarization.[37] Organic ferroelectrics are suitable candidates to calculate the electronic polarization; in ref. 34, Kobayashi *et al.* performed polarization experiments on the organic molecular crystal tetrathiafulvalene *p*-chloranil (TTF-CA) that showed giant electronic polarization below 81 K where donor TTF donates one electron to acceptor CA, while at the same time a small shift in upward direction was observed which contributes to the ionic polarization. The electronic polarization may be additive or subtractive to the ionic polarization, depending on the electron transfer contribution and hybridization of electronic orbitals. In extreme conditions there is always a possibility for electrons to dominate total polarization. In the case of $BaTiO_3$ electronic polarization is in the same direction as that of ionic polarization; however, in $HoMn_2O_5$ crystals polarization is driven by spin, and electronic polarization contributes in the opposite direction to ionic polarization.[38] It might be also possible that the switching of polarization is due to generation of large amount of photo-induced charge carriers in the direction of polarization which significantly redistribute the electron density. To check such photoconductive effects, the samples were poled for a long time to neutralize domain pinning, compensate the trapped charges at grain boundaries, and fill the oxygen vacancies with free charge carriers. The poling was quite effective, which significantly reduced the enhancement of polarization under illumination; however, the effect was still significant since polarization increase remained at nearly 50-150 % under light, which indicates intrinsic contribution of electronic polarization. [39,40,41,42,43] These results also suggest that there is a possibilities that the shift current associated with asymmetry between valence and conduction band spatial distributions under illumination may also significantly affect the displacement current and overall polarization.[44]

Fig.4 shows displacement current of unpoled and poled PLBZT under dark, white light, and monochromatic light at various frequencies. Sharp peaks in displacement currents were observed in both conditions which become more prominent under illumination. These results suggest that irrespective of resistive contribution in polarization/displacement current, a significant amount of displacement current also switches under illumination.

To further strengthen the claims for enhancement of polarization under illumination, a PUND (Positive Up and Negative Down) measurement has been carried out to determine the logic states for ferroelectric memory elements based on a series of five pulses. PUND measurements reliably discriminate between charge injection and true ferroelectric displacement currents; since P is actually measured by integrating charge in ferroelectric polarization measurements, this is important. PUND measurements were performed on both unpoled and poled ceramics at 80 kV/cm electric field with a pulse width of 10 ms and 100 ms and a delay time of 1000 ms, as shown in Fig. 5(a-d). Switchable polarization can be calculated as follows:

dP = P$^*$ - P$^\wedge$ = where P$^*$ = (switchable polarization + non-switchable polarization) and P$^\wedge$ = (non-switchable polarization); dP$_r$ = P$^*_r$ - P$^\wedge_r$ where P$^*_r$ = (switchable remanent polarization + non-switchable remanent polarization) and P$^\wedge_r$ = (non-switchable remanent polarization). PUND analysis also indicates almost 50-150 % enhancement in remanent polarization under illuminated conditions, confirming the earlier P-E loops and displacement current results.

To check the reproducibility of the polarization switching phenomenon, a series of P-E measurements on unpoled and poled samples have been carried out for several days and within intervals of a few minutes. The systems showed consistently giant saturated polarization switching for days, as shown in a bar diagram in Fig. 6 (a-d). The unpoled and poled samples show similar trends for polarization switching under light. The fast switching and recovery (under darkness) of polarization under illumination are the sole criteria for use as electrical write and optical read nonvolatile-memory elements. Fast polarization switching and its recovery within minutes were obtained over a long period of time. The polarization under light consistently reverts back to normal polarization states as soon as we switch off the light source. The effects of white light, monochromatic light and probe frequencies on polarization switching have been made within the permissible limit of experimental errors. These studies revive attention towards the device application of switching of photo-ferroelectric polarization.

### 3.3. Leakage current, photo-current, and Fatigue properties

We have chosen white light for long-time exposure experiments such as leakage current, open circuit voltage ($V_{OCV}$) and short circuit current ($I_{SCC}$), ON and OFF states of photo-current, and PUND-based polarization fatigue testing, as shown in Fig. 7(a-d). The leakage current measured in dark and white light shows almost symmetrical leakage current behavior until ± 80kV/cm E-field. The average dark current is about 5 µA at -/+ 80 kV/cm, whereas the level of leakage current increases drastically -- almost one order of magnitude ~ 0.3 µA -- under the illumination of white light (Fig. 7a). This enhancement in leakage current is due to optically active charge carriers which gain energy under light illumination, jump from valence band to conduction band and cross the Schottky electrode-ferroelectric barrier; however, these charge carriers are significantly suppressed under E-poling to fill the traps, voids, oxygen vacancies, and neutralization of space charge across the outer surface of grain boundaries.[41] E-poling effectively reduces the trapped charges across grains and grain boundaries and compensates defect density.[41] The inset in Fig. 7 (a) shows the leakage current behavior after E-poling and under illumination that suggest significant reduction of photo-charge carriers after poling. Interestingly, PLBZT also shows ferroelectric photovoltaic effects with a small amount of $I_{SCC}$ ~ 0.6 nA and development of a open circuit voltage of almost $V_{OCV}$ ~ 2 V (Fig. 7b). It indicates natural development of a p-n junction across insulating domains and conducting domain walls. Ferroelectric photovoltaics have several advantages over conventional silicon solar cells: They can provide $V_{OCV}$ beyond the bandgap of materials, natural p-n junctions, and switching of photocurrents based on polarization direction; however, the major drawback is low $I_{SCC}$, which make them weak candidates as potential solar devices. Yet a large number of scientific groups are working on hybrid organic-inorganic solar cells and perovskite solar cells to improve efficiency.[45,46,47] It can be easily seen from the inset of Fig. 7 (b) that the direction of photo-current changes with change in direction of polarization. It indicates that the direction of polarization governs the photo-currents; however, these effects were not symmetrical due to the large thickness of our electro-ceramics, which does not allow light to reach the bottom electrode-electrolyte interface or for diffusion of photo-charge carriers to the opposite electrode. The internal electric field due to domains drives photo-generated electrons and holes in opposite directions.[48,49] Semiconductor ferroelectrics are known for the generation of photo-currents under illumination due to change in Schottky barrier height across the electrode-electrolyte interface and develop open circuit voltages due to the difference in chemical potential between the different charge carriers.[50,51]

Switchable ON and OFF photo-current characteristics of PLBZT were recorded under light and dark conditions with a bias voltage of ±100V (thickness around 500 µm). A moderate ON/OFF ratio ~ 6:1 of photo-current was observed with sharp switching and slow decay over long period of time (Fig. 7 (c). Typical switching and decay times were observed to be ca. 34 seconds (+/- 2s) and 43 s (+/- 2 s), respectively, within the resolution of current-voltage source meter (1s). A mathematical model has been used to calculate the time constants of the material under light; we discuss this next. To measure the lifetime of devices and their performance under large electric field, the PUND based fatigue characteristics have been measured with 1 Hz frequency and 20 kV/cm electric field. These results show interesting features in remanent polarization under dark and light conditions. The switchable remanent polarization vs. cumulative time is shown in Fig. 7 (d). The fatigue results suggest negligible (~20%) degradation of remanent polarization (dP) after $10^8$ cycles of large electric field under dark conditions; however, interestingly, dP improves under illumination for a similar period of time. Note that most of the PZT ferroelectrics showed nearly 40 to 50 % polarization deterioration after $10^9$ cycles of electric field. These results suggest that light illumination not only switches the polarization but also improves under constant illumination.

## 3.4 Photo-response time study

The photocurrent ON/OFF conditions were fitted to a mathematical model based on the movement of fast and slow charge carriers and their lifetimes, as described below. To determine the lifetimes of slow and fast charge carriers during the growth and decay process, the ON/OFF currents, were fitted to the model below.[52,53,54] Equation 1 is used to fit the experimental data of decay process.

$$I(t) = I_o + I_{fm} \exp\left(\frac{-t}{\tau_{fm}}\right) + I_{sm} \exp\left(\frac{-t}{\tau_{sm}}\right) \quad \ldots\ldots(1)$$

During the photo-current decay process, an exponential decay equation has been used to calculate the lifetime of charge carriers, where t is time; $\tau_{fm}$ and $\tau_{sm}$ are lifetimes of the fast and slow recombination charge carriers; $I_{fm}$ and $I_{sm}$ are fast and slow photo-charge carrier currents; $I_0$, offset current. A clear view of experimental data of growth and decay process of photocurrents are given in Fig. 8 (a) where as these experimental data fitted well with the decay (Eq.1) and growth (Eq.2) model of photo-generated charge carriers, as shown in Fig. 8 (b). The solution gives a lifetime of fast recombining charge carriers ($\tau_{fm}$) in the range 11 s, whereas the lifetime of remaining slow recombining charge carriers ($\tau_{sm}$) nearly of 32 (±2) s reaches saturation (dark current).

The mathematical model used during the growth process is mentioned below.

$$I(t) = I_o + I_{fm}\left[1 - \exp\left(\frac{-t}{\tau_{fm}}\right)\right] + I_{sm}\left[1 - \exp\left(\frac{-t}{\tau_{sm}}\right)\right] \quad \ldots\ldots(2)$$

The solution gives a lifetime of fast photo charge carriers in the range 2 s with lifetime of remaining slow photo charge carriers of 32 (±2) during growth process. These data suggests that during the growth and decay process, photo induced charge carriers switched from initial state (dark current) to final state within 40 to 50 seconds and vice versa; these response times with proper microelectronics may provide photo-detectors in the visible and near UV range.

## 4. Conclusions

In summary, we have successfully synthesized an optically active single-phase PLBZT electro-ceramic having tetragonal crystal structure at room temperature which undergoes a ferroelectric to paraelectric phase transition near 380 $^0$C. A moderate change in dielectric constant (~3%) under light was observed at room temperature for low probe frequency; however, a significant decrease in the magnitude of dielectric constant was observed near the ferroelectric phase transition under light. This material showed almost 200-300 % increase in polarization under illumination of monochromatic light; the effect of light decreased to 50-150% after E-poling for a long time. We have explained these phenomena due to development of electronic polarization and alignment of ionic polarization in the direction of applied E-field under light. Apart from giant enhancement in polarization, it illustrates significant photovoltaic effects with weak $I_{SCC}$ (~ nA) and moderate $V_{OCC}$ (~ 2V) and large ON/OFF photocurrents with a ratio of 6:1 under illumination of weak light. Interestingly, it has also been found that under PUND based fatigue test, PLBZT displayed remarkably improved polarization after $10^9$ cycles with large electric field. These properties of newly discovered optically active ferroelectrics suggest that they might be potential candidates for future electrical-write and optical-read nonvolatile memory elements. They should also be tested further as suitable candidate for ferro-photovoltaic devices, near ultra-violet sensors and energy harvesters.

**Acknowledgement:**

AK acknowledges the CSIR-MIST (PSC-0111) project for their financial assistance. Hitesh Borkar would like to acknowledge the UGC (SRF) to provide fellowship to carry out Ph. D program. Authors would like to thank DNPL, Dr. Ms. Ranjana Mehrotra and Dr. Sanjay Yadav, for their constant encouragement.

**Figure captions**

**Fig. 1** XRD patterns of PLBZT electro-ceramics; solid line illustrates Rietveld refinement data. Inset shows HRTEM images of (100) lattice planes matched with XRD data.

**Fig 2** Dielectric constant and tangent loss spectra as functions of temperatures and frequencies (a,b), temperature dependent comparative dielectric data at 1kHz and 5 kHz (c,d), and loss tangent data (e,f) in dark or illumination of laser light with energy density 30 mW/cm$^2$.

**Fig. 3** Enhancement of ferroelectric polarization under dark, white light and laser light for 2 and 5 Hz (a,b) for unpoled and (c,d) E-poled electro-ceramics.

**Fig. 4** The displacement current under dark, white light and laser light for 2 and 5 Hz (a,b) for unpoled and (c,d) E-poled electro-ceramics.

**Fig. 5** Enhancement of ferroelectric polarization using PUND measurements under dark, white light and laser light for 10 ms and 100 ms pulse width and 1 second pulse delay time, (a,b) for unpoled and (c,d) E-poled electro-ceramics.

**Fig. 6** (a,b) Electrical polarization switching behavior of PLBZT at 1Hz and 2 Hz with time interval of days and (c,d) with time interval of 3 minutes. Encircled data indicate the switching statistics of unpoled electro-ceramics

**Fig. 7** (a) Leakage current measurements in the dark and under white light for unpoled ceramics (inset shows I-V data of poled sample); (b) Current-voltage measurement with solar simulator (with 70 mW/cm$^2$) for the photovoltaic analysis; (c) ON/OFF switching mechanism for photo-generated charge carriers; (d) Fatigue characteristics for unpoled ceramics with and without white light.

**Fig. 8** (a) Clear-view of ON/OFF switching behavior of photocurrents, (b) double exponential fitting for growth and decay of photocurrents at 180s period.

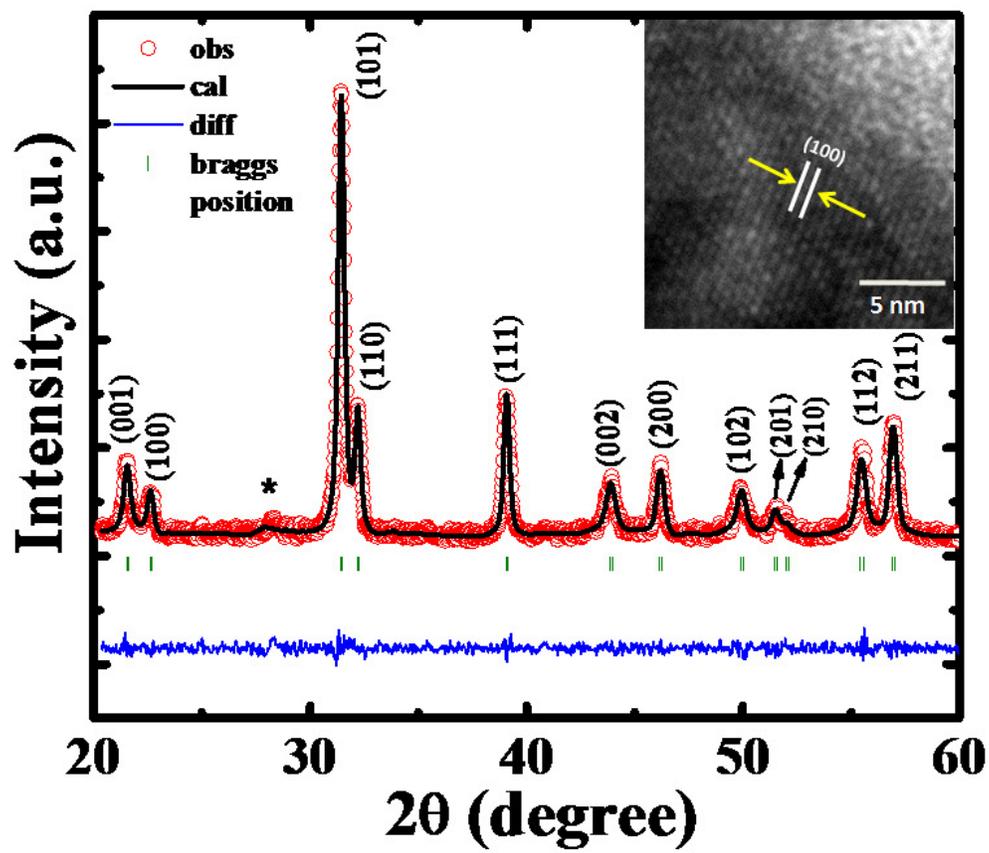

Fig. 1

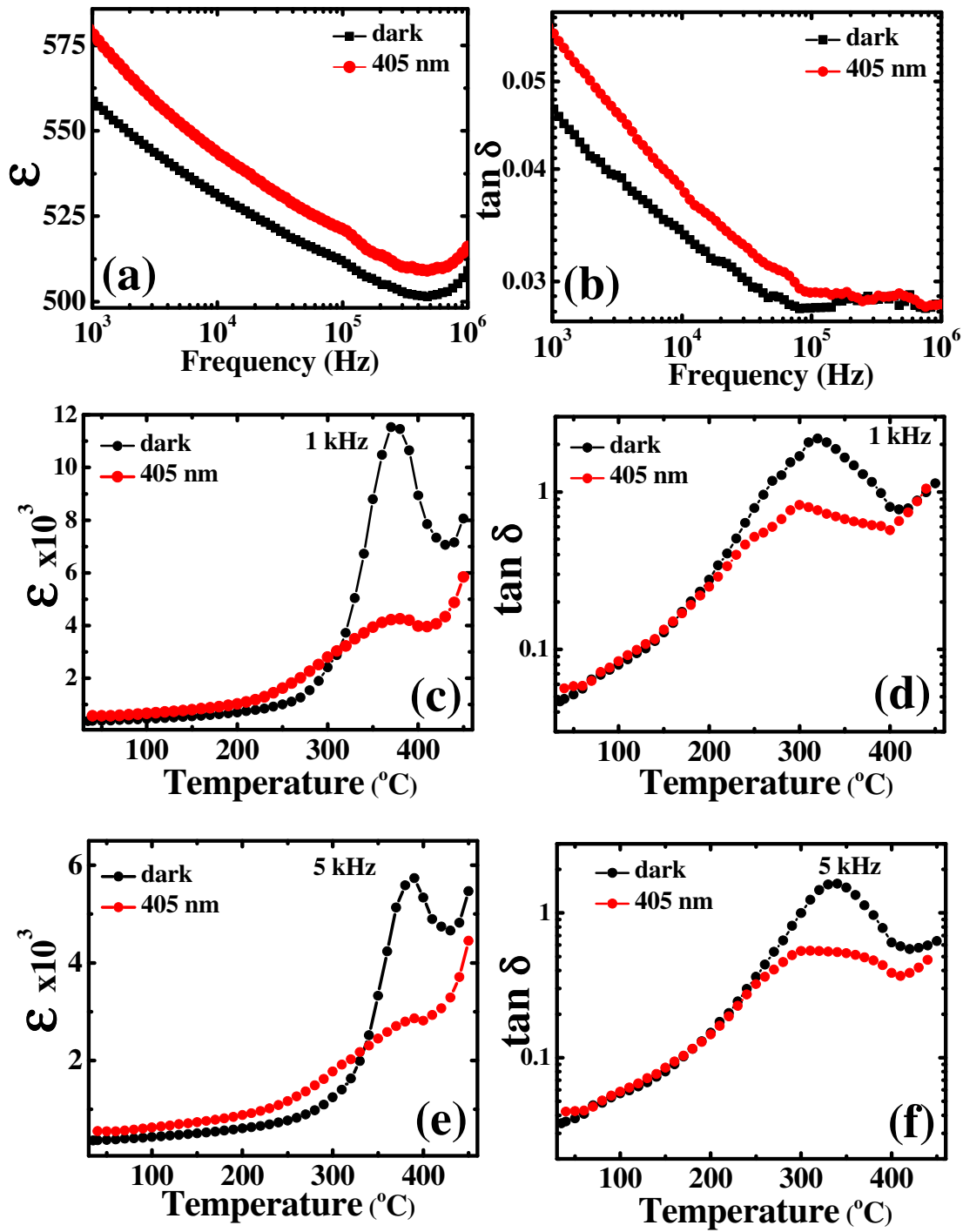

**Fig. 2**

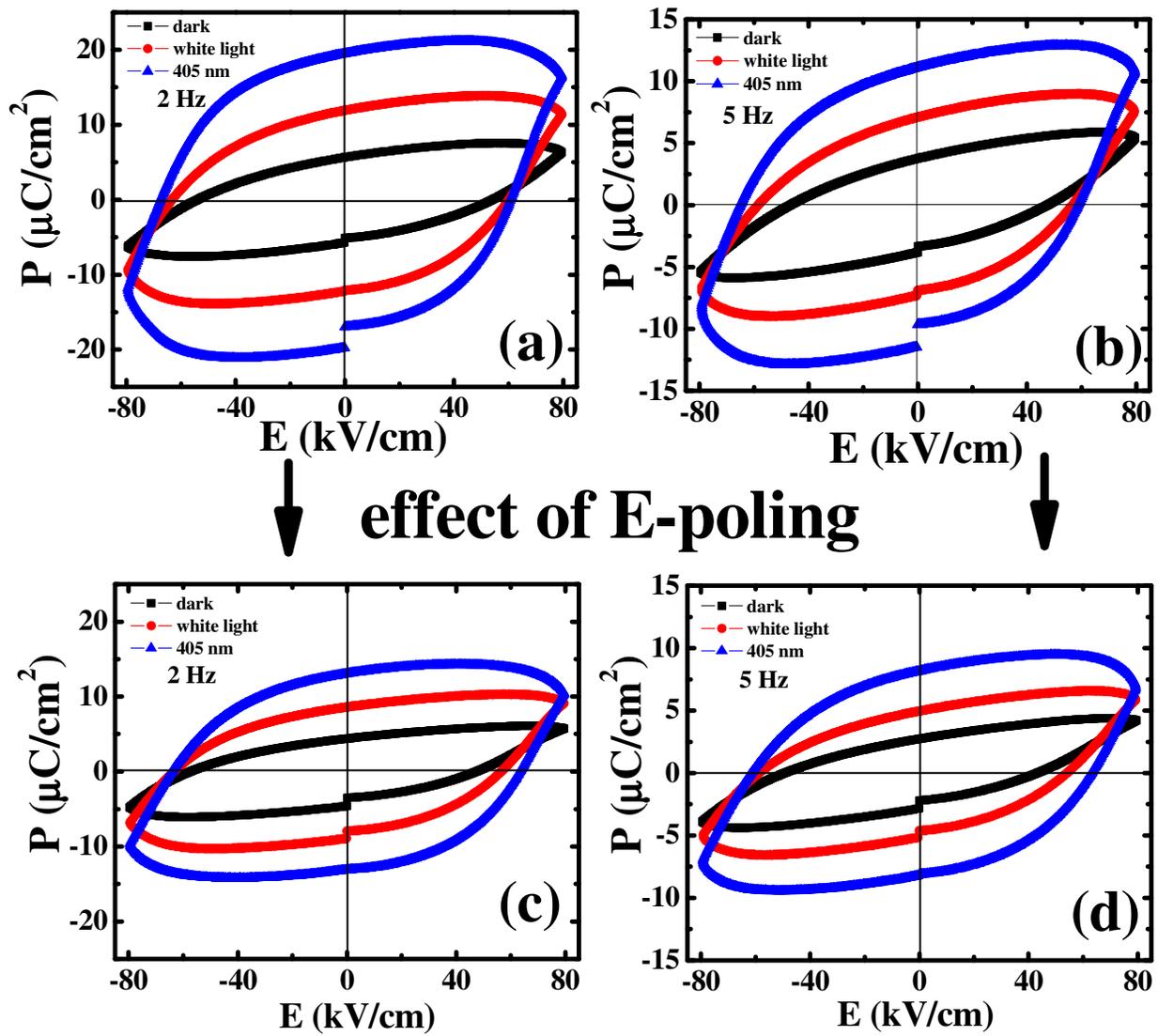

**Fig. 3**

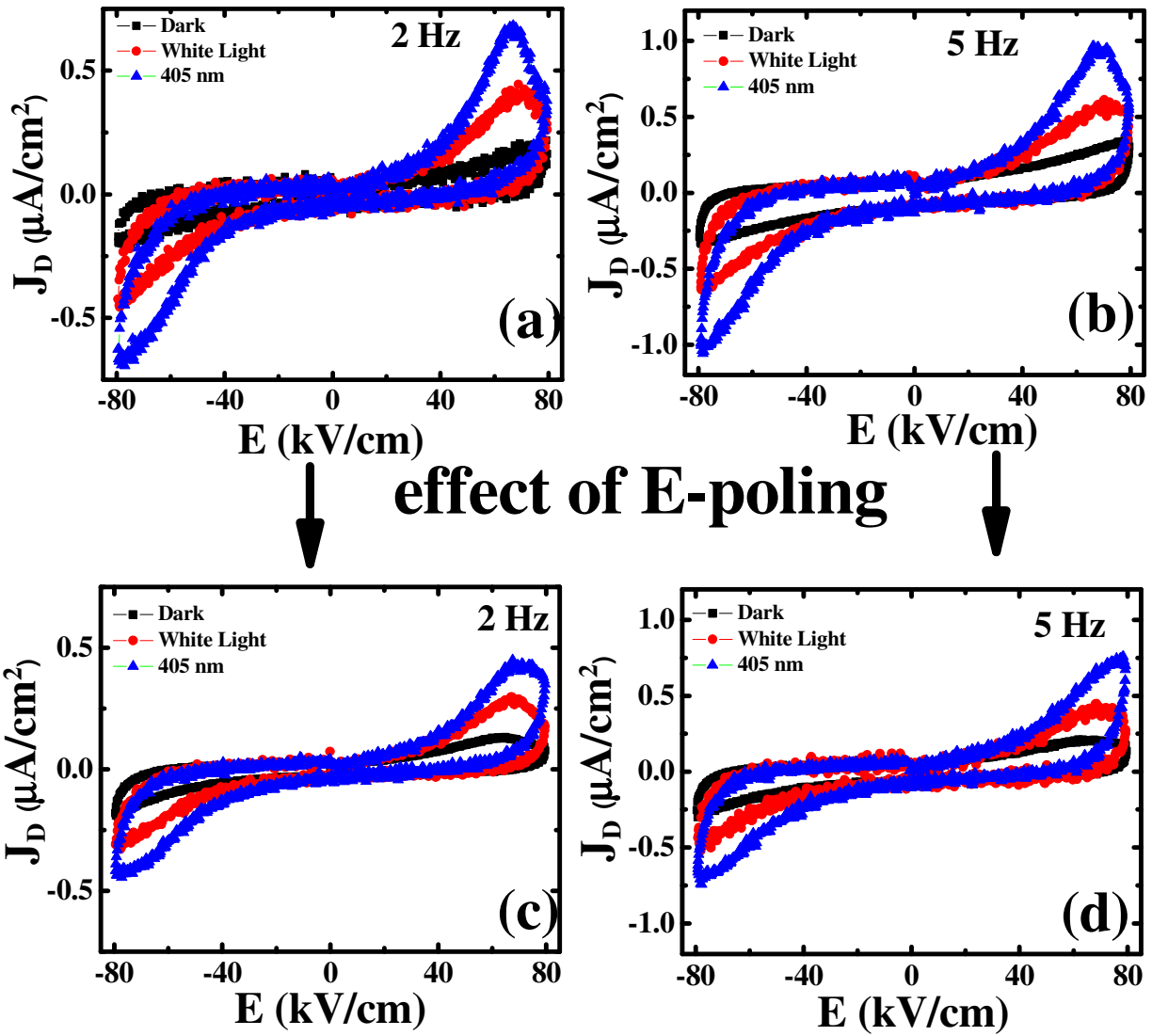

**Fig. 4**

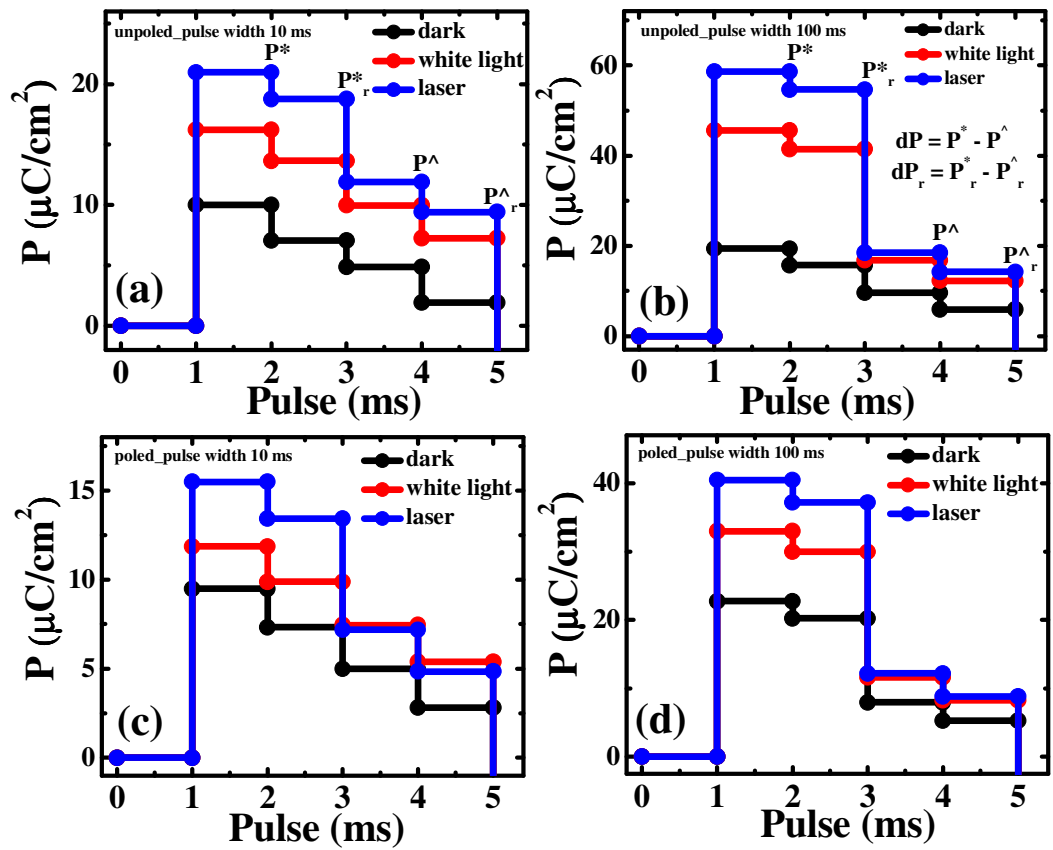

**Fig. 5**

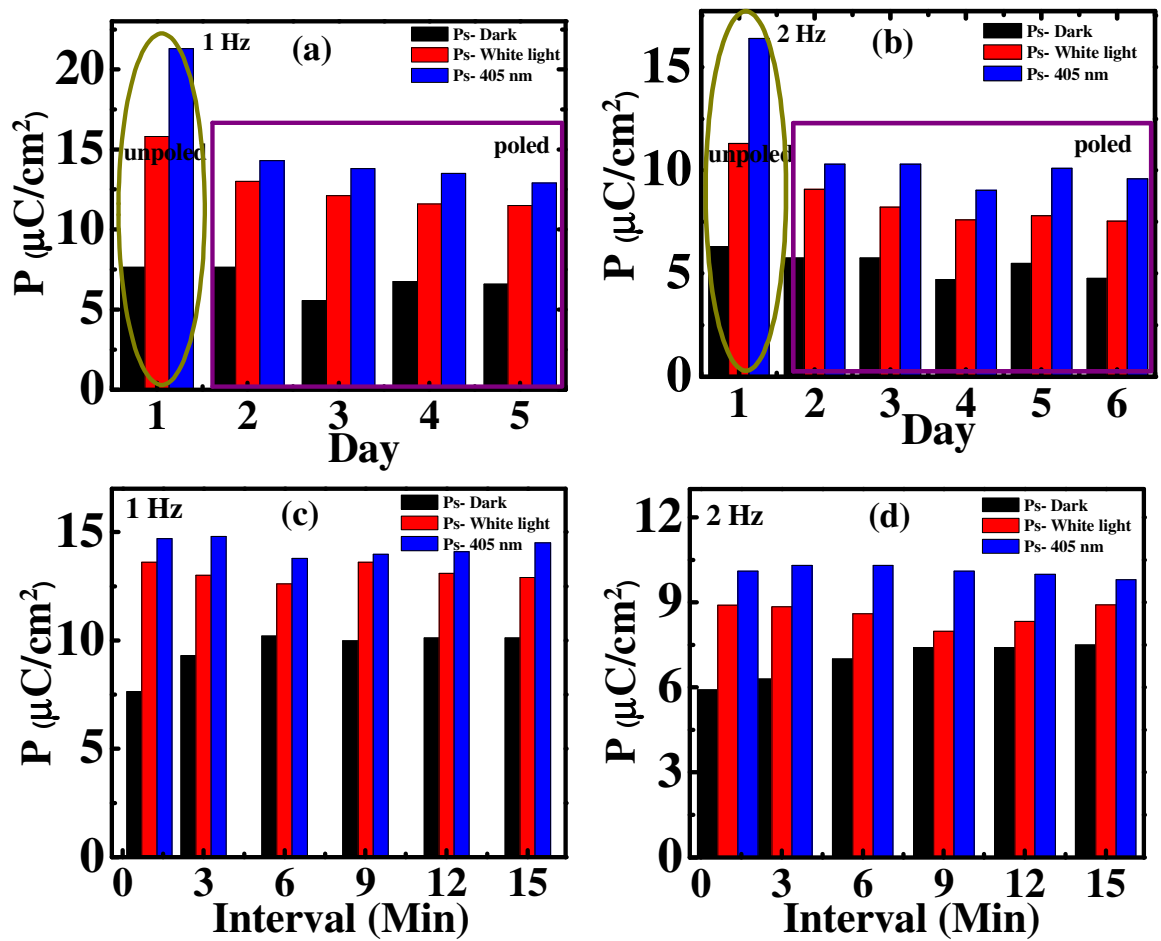

Fig. 6

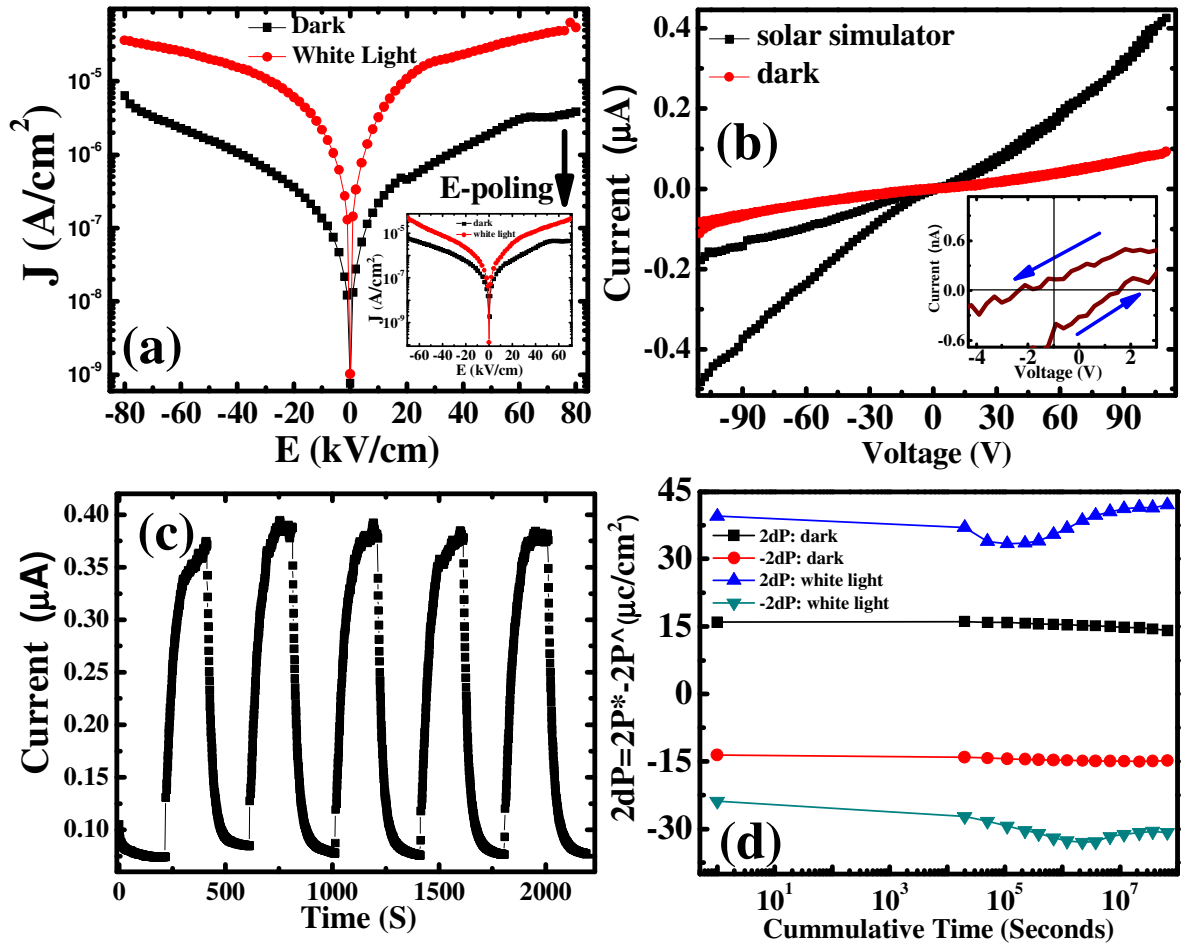

**Fig. 7**

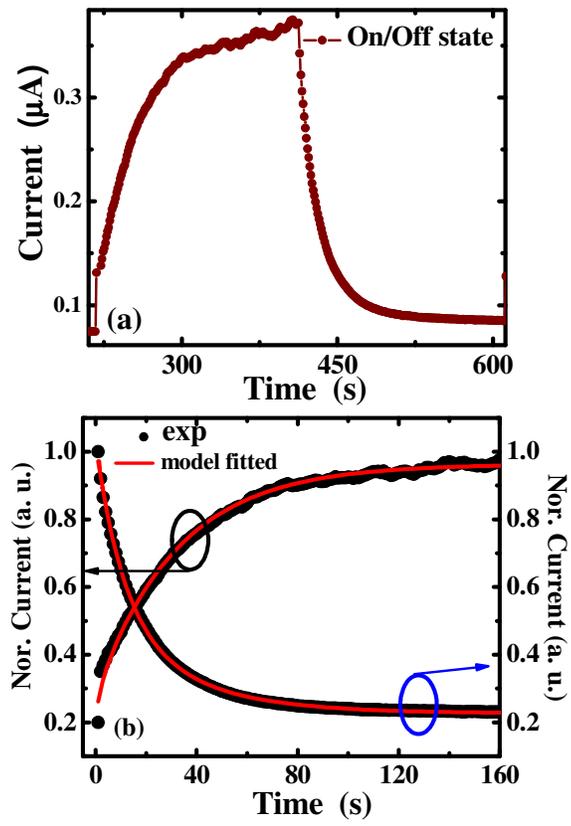

**Fig. 8**